\pdfoutput=1
\documentclass[manuscript]{aastex}
\usepackage{natbib}
\usepackage{graphicx}
\usepackage{subfigure}

\slugcomment{Received ........; Accepted .........}

\shorttitle{Twisting motions in sunspot penumbrae.}

\shortauthors{Bharti et al.}

\begin{document}

\title{Waves as the source of apparent twisting motions in sunspot penumbrae}

\author{L. Bharti$^{1}$, R.H. Cameron$^{1}$, M. Rempel$^2$, J. Hirzberger$^{1}$ and S. K. Solanki$^{1,3}$ }
\affil{1. Max-Planck-Institut f\"ur Sonnensystemforschung, Max-Planck-Str. 2, 37191 Katlenburg-Lindau, Germany}
\affil{2. High Altitude Observatory, NCAR, P.O. Box 3000, Boulder, CO 80307, USA}
\affil{3. School of Space Research, Kyung Hee University, Yongin, Gyeonggi Do, 446-701, Korea}

\email{bharti@mps.mpg.de}

\begin{abstract}
The motion of dark striations across bright filaments in a sunspot penumbra has become an important new diagnostic of convective gas flows in penumbral filaments. The nature of these striations has, however, remained unclear. Here we
present an analysis of small scale motions in penumbral filaments in both simulations
and observations. The simulations, when viewed from above, show fine structure with dark lanes running
outwards from the dark core of the penumbral filaments. The dark lanes either occur preferentially on one side
or alternate between both sides of the filament. We identify this fine structure with transverse (kink) oscillations of the filament,
corresponding to a sideways swaying of the filament. These oscillations have periods in the range of 5-7 min and propagate outward and downward along the filament. Similar features are found in observed G-band intensity time series of penumbral filaments in a sunspot located near disk center obtained by the
Broadband Filter Imager (BFI) on board {\it Hinode}.
We also find that some filaments show dark striations moving to both sides of
the filaments. Based on the agreement between simulations and observations we conclude that the motions of these striations are caused by transverse oscillations of the underlying bright filaments.

\end{abstract}


\section{Introduction}

High resolution Hinode and Swedish Solar Telescope observations of sunspots located away from disk center have revealed fine structure
in penumbral filaments: filaments which are nearly perpendicular to the solar disk radius
vector display a "twisting motion" indicated by dark striations moving across the filaments (i.e. perpendicular to the filament's axis) always directed from the limb-side to the center-side of the filament \citep{ichimoto_07}.
These striations can be used as tracers of the flow, and have observationally established the presence of overturning
convection in the filaments \citep{zakharov08,bharti10a}, an idea which is consistent with theory and recent
simulations \citep{heinemann07,Rempel2009b,Rempel2009}.
The direct measurement of the velocity is more difficult: \cite{almeida07} found a local correlation between upflows
and bright  structures as well as between downflows and dark structures in a penumbra. Such correlations
are suggestive of convective energy transport in the penumbra.
The clear signal of an upflow along the central axis of a bright filament
has been reported by \cite{franz09}, \cite{bellot10} and  \cite{ichimoto10}, downflows at its sides are more challenging to observe. Line syntheses
from sunspot simulations (cf. \cite{bharti11}) suggest that these downflows are partly hidden in observations due to both limited spatial resolution, and
the fact that commonly used lines form above the heights where the downflows are
strongest.  Moreover, the Evershed flow affects line of sight velocities, thus
the detection of downflows at edges of filaments also depends on the
location of the sunspot on the solar disk. Recently \cite{Joshi11} and \cite{Scharmer11} found
downflows in dark regions at the edges of the penumbral filaments in the C I 5380\,\AA~ line formed deep in the photosphere. The results of these publications support the prediction made by \cite{bharti11}. \cite{Scharmer12} reported that downflows at the edges of filaments detected in the C I 5380\,\AA~ line (\cite{Scharmer11}) are also present in the wing of Fe I 6301.5 \AA~ with reduced amplitude.

The striations have the advantage that they potentially allow horizontal motions to be followed, if detected close to disk center, which would make them of unique diagnostic value for the velocity field we have.
In previous studies by \cite{ichimoto_07}, \cite{zakharov08}, \cite{spruit10} and \cite{bharti10a}, the apparent twist
of the striations was analyzed only in sunspots away from the disk center where the twisting motions is seen
in filaments perpendicular to the line of symmetry (i.e. in filaments directed parallel to
the nearest portion of the limb).  The twists are always directed toward the center-side. This was interpreted as a geometrical effect by
\cite{zakharov08} -- these "twisting" motions were exclusively interpreted in terms of overturning convective flows
perpendicular to the filament's major axis \citep{ichimoto_07,zakharov08,scharmer09,spruit10}.
\cite{spruit10} proposed that these striations originate from the "corrugation of the boundary between an overturning convective flow inside the filament and the magnetic field wrapping around it". Based on their modeling results, they also argue that the striations are not compatible with a horizontal field along the axis of filaments in excess of 300 G. However, it is not certain if they really trace convective flows or not. This is an important point to establish, since the motions of the striations have been employed by \cite{zakharov08} to conclude that convective motions transport sufficient energy to explain the brightness of the penumbra (cf. \cite{bharti10a}) . In this paper we use the "realistic" numerical radiative MHD  simulations of \citep{Rempel2009b,Rempel2009,Rempel2011} to investigate the causes of the observed brightness striations.
In addition, we  analyze such striations in penumbral structures
observed at disk center and compare their properties with those found in the numerical simulations.

The paper is organized as follows: in Section 2 we describe the numerical simulation and
present an analysis of the fine structure we find there, in Section 3 we describe the
observations and compare the simulations with the observations. We then
present our conclusions in Section 4.

\section{Numerical simulations}
The simulations analyzed here were carried out with the MURaM code \citep{Voegler05}.
The code includes the effects of partial ionization on the equation of state, and non-grey
radiative transfer. For details of the code and the equations see \cite{Voegler05} and for recent
modifications, essential for the sunspot simulations presented here, see \cite{Rempel2009}.
This code has been used extensively to treat problems both in the quiet Sun
\citep{Keller04,Khomenko05,Voegler07,Pietarila_Graham09,Graham10, Danilovic10b, Danilovic10a} as well as  flux concentrations reaching from pores to entire active
regions \citep{Cameron07,Cheung08, Cheung10,Yelle_Chaouche09,Schuessler06,Rempel2009b,Rempel2009,bharti10b,Rempel2010,Rempel2011}.

Here we present results from two different simulation runs. The first simulation uses a setup in
'slab' geometry, in which only a narrow slice through the center of a sunspot is simulated. The geometry
and size of this simulation make it ideal for studying the detailed 3 dimensional evolution of
individual penumbral filaments, albeit in a somewhat artificial geometry.
The second simulation uses a setup with a pair of opposite polarity sunspots, leading to more extended
penumbrae with a more realistic geometry.

\subsection{Slab Geometry}
A snapshot from the sunspot simulation in slab geometry described by \cite{Rempel2009} was used as the initial condition
for the calculations presented here. The simulation domain is periodic in both horizontal ($x$ and $y$) directions,
with dimensions of 4.6~Mm $\times$ 36.864~Mm, and has a dimension of 6.144 Mm in the vertical ($z$) direction.
The average value of the $\tau_{\mathrm{Ross}}=1$ ( $\tau_{\mathrm{Ross}}=1$ levels refer to optical depth computed from the grey opacities (which is Rossland mean opacity, an "average" opacity such that if we assume that the opacity at all frequencies is this average.))height in the quiet Sun is used to define $z=0$, and $z$ is defined
to be positive above this height and negative in the interior of the Sun.
The vertical boundary conditions are unchanged from \cite{Rempel2009}: the top boundary is closed and the magnetic field above it
is assumed to be potential. The bottom boundary is open as described in \cite{Voegler05}.

From this initial condition, the simulation was continued for 133 minutes of solar time,
with snapshots saved every 34.5s.

\subsubsection{Simulation analysis}
\begin{figure}
\includegraphics{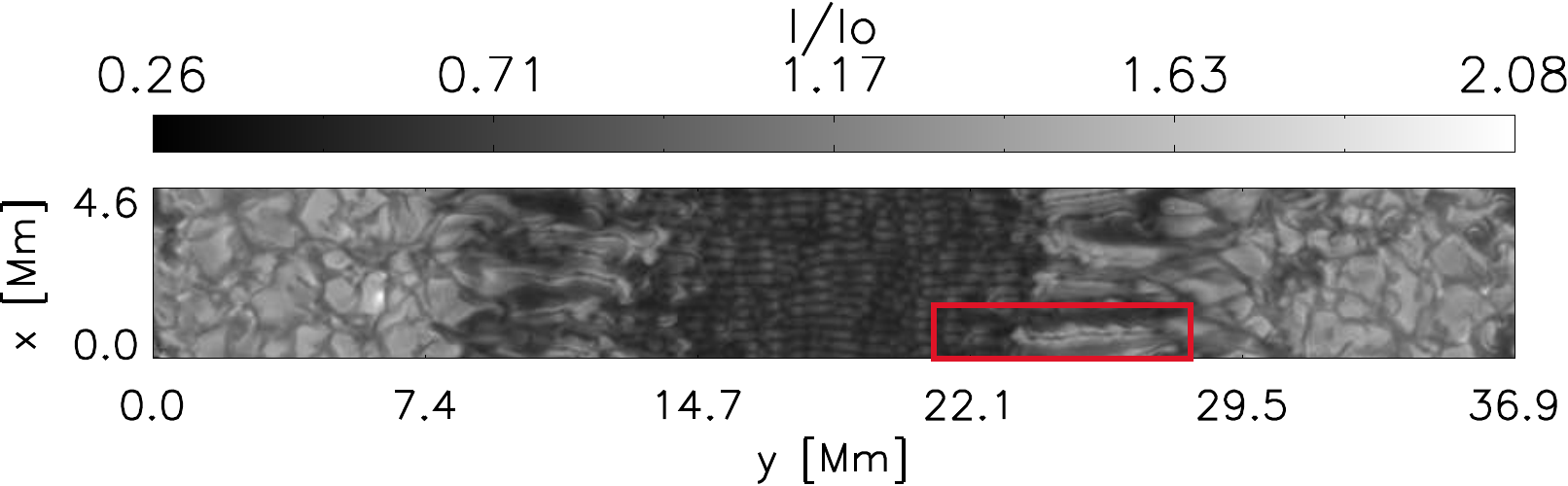}
\caption{Bolometic intensity from the simulation in slab geometry at $t=71.9$min. The red box indicates the filament
we chose for further study.}
\label{fig:sim_ref}
\end{figure}

\begin{figure}
\includegraphics[width=14cm]{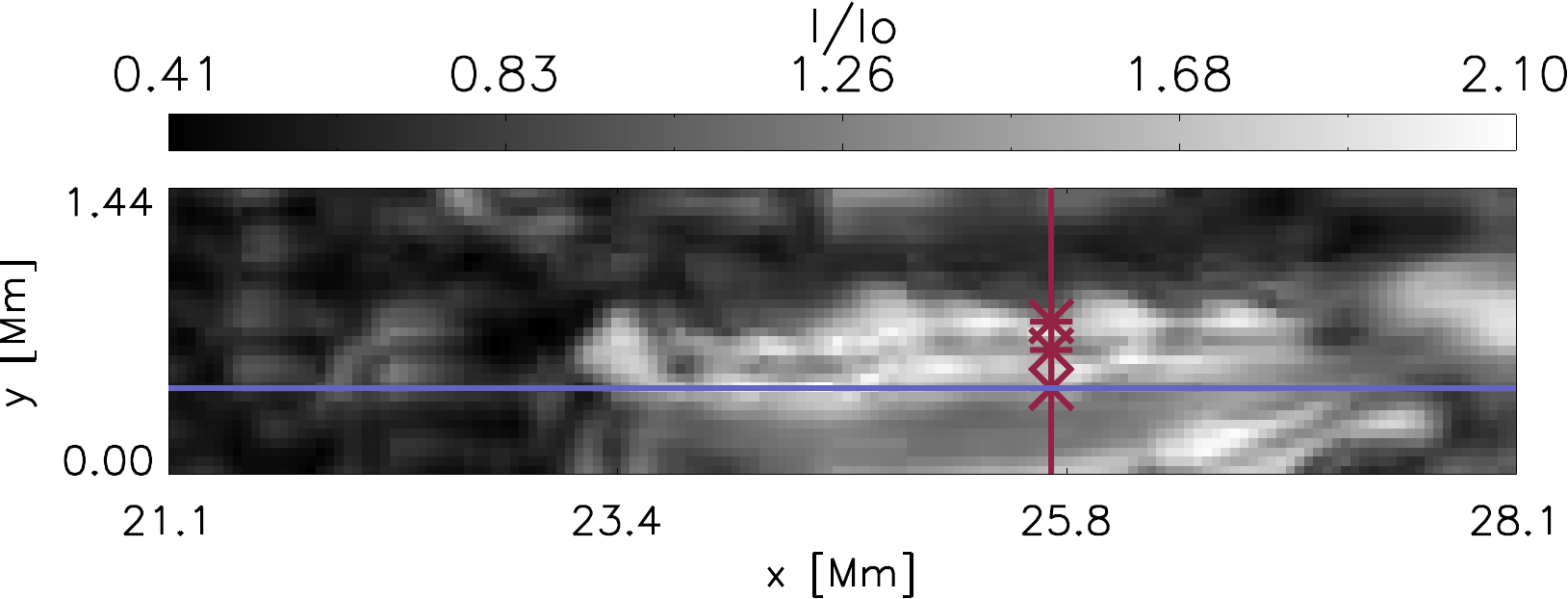}
\vspace{0.5cm}
\includegraphics[width=14cm]{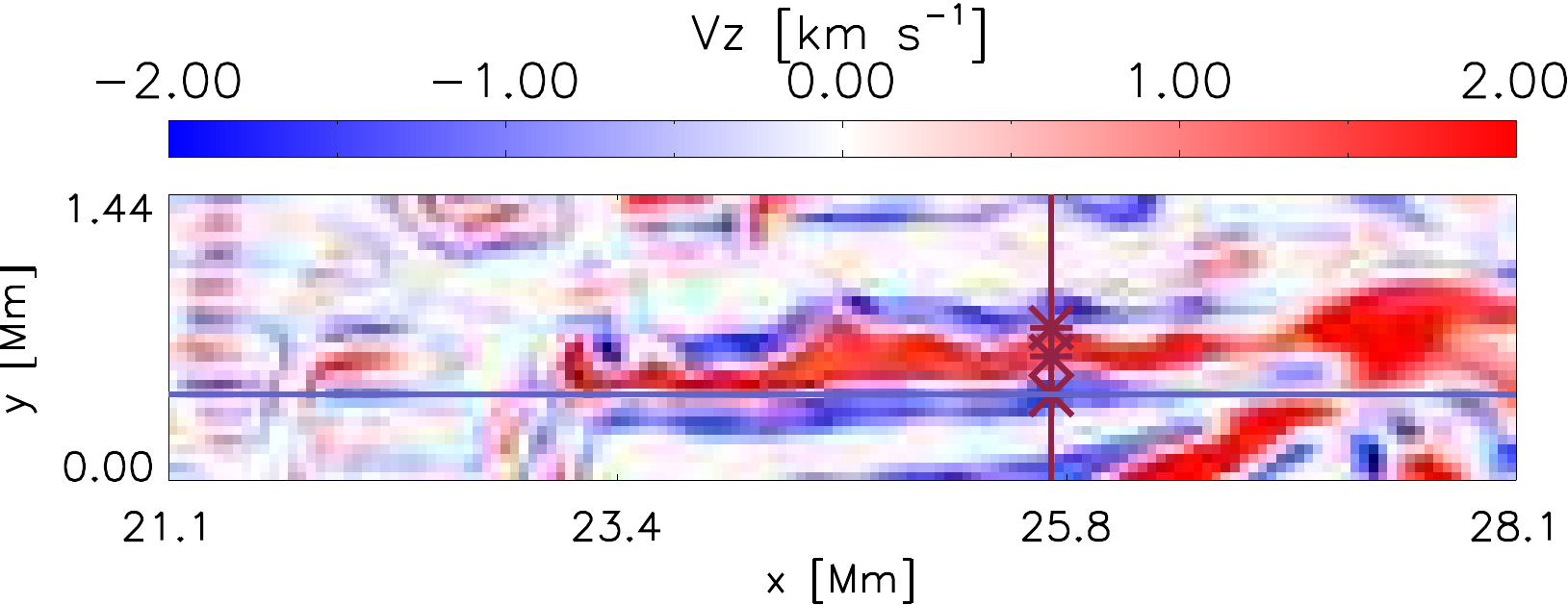}
\includegraphics[width=14cm]{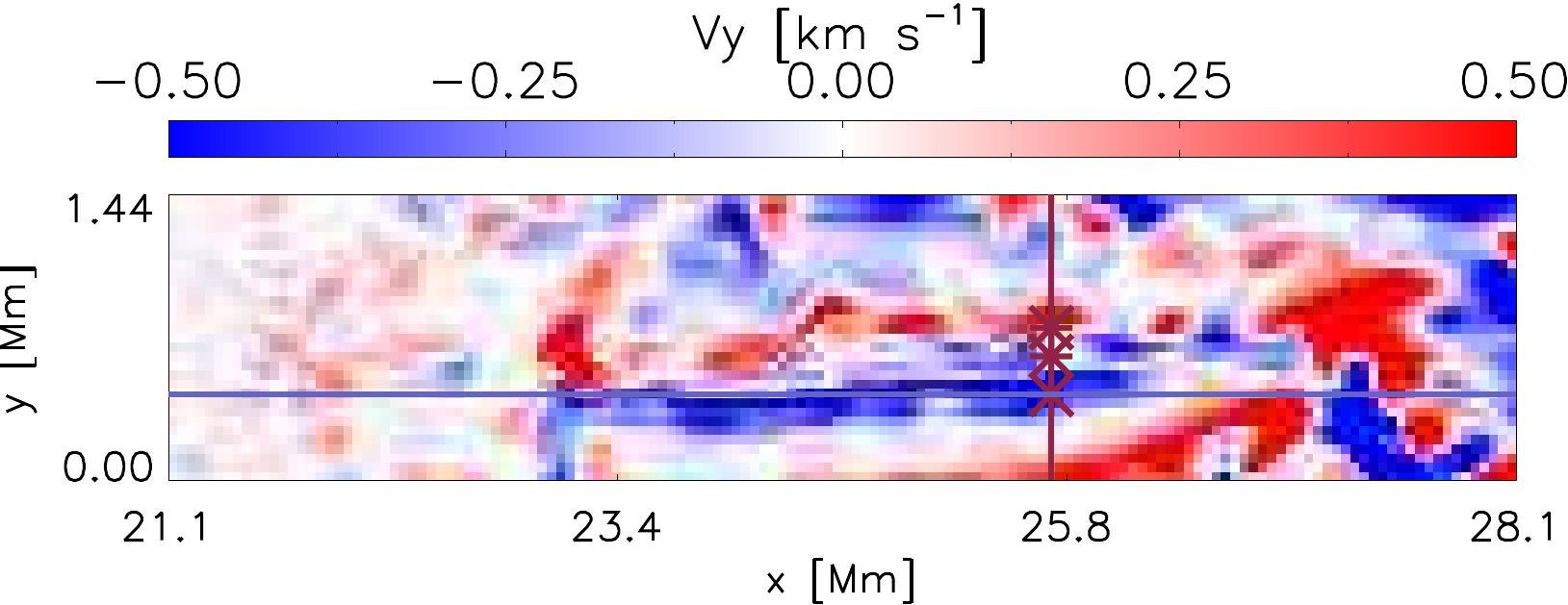}
\caption{Intensity map and horizontal cuts through the simulated filament in the red box in
Fig~\ref{fig:sim_ref}.
From top to bottom, the panels show: the bolometric intensity; the component of the velocity in the
$y$ direction saturated at $-500$~m/s (red) and $+500$~m/s (blue) at a height
$z=-384$~km below the average height of
$\tau_{\mathrm Ross}=1$ in the quiet Sun ; and the component of the velocity in the $z$ direction
saturated at $-2$~km/s (red) and $+2$~km/s (blue) at the same height. The red line at   $x=25.6$~Mm,
the red crosses and the blue line at  $y=0.51$ Mm show the locations at which slices were taken
for further analysis (black in lower frames).}
\label{fig:I_times}
\end{figure}

\begin{figure}
\includegraphics{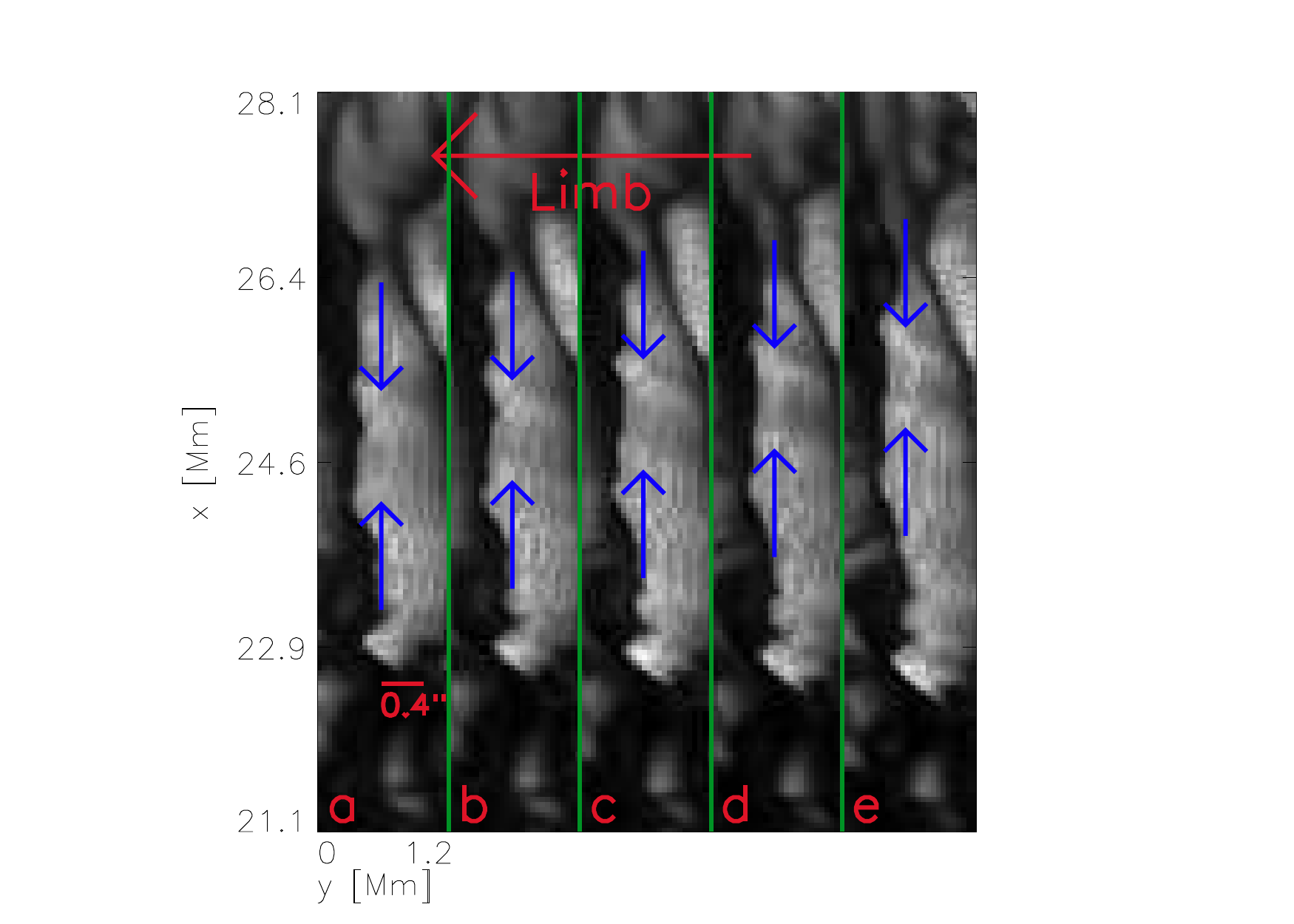}
\caption{Snapshots of the simulated penumbral filament seen at an angle of
$50^{\circ}$ from disk centre at times (a) $t=57$~min, (b) 60~min, (c)
63 min, (d) 66~min and (e) 69~min. The umbra is situated at the bottom of the figure and the quiet-Sun above the top of the figure. At this viewing angle,
inclined striations,
indicated by the blue arrows, can be seen propagating away from the umbra.
}
\label{fig:I_inc}
\end{figure}

The bolometric intensity   of the entire simulation domain viewed from above at $t=71.9$~min
is shown in Fig.~\ref{fig:sim_ref}. Several penumbral filaments with central dark cores can be seen.
In the following we consider the filament in the lower right region of the penumbra ($21.1 \le y \le 28.1$~Mm
and $0 \le y \le 1.44$~Mm). A blow up of this region is shown in Fig.~\ref{fig:I_times}.
There  we see that, at the height of the cut $z=-384$~km (i.e. 384~km below the average $\tau_{\mathrm{Ross}}=1$
height of the quiet Sun) there is a continuous upflow along the central part of the filament and downflows along the borders.
This cut lies below the local $\tau_{\mathrm{Ross}}=1$ surface, so that the shown up and downflows are not directly observable.
The filament exhibits fine structure in the form of 'wiggles'
in the bolometric intensity as well as horizontal and vertical velocities along its entire length.
The wavelength in the $x$ direction of the 'wiggles', at this fixed time, is approximately 700~km.
Figure~\ref{fig:I_inc} shows that, when viewed at an angle, inclined striations which propagate away
from the umbra appear, somewhat similar to what is seen in the high resolution observations by \cite{ichimoto_07}, \cite{zakharov08}, \cite{spruit10} and \cite{bharti10a}.

In order to study the time evolution of the fine structure, we first focus on cuts
across the filament at $x=25.7$~Mm. The time evolution of the cuts for various quantities
are shown in Fig.~\ref{fig:I_time_slice}. The range
of the intensity image has been restricted in order to better reveal the fine structure
between $t=50$ and $t=90$~min. Between $t=0$ and $t=30$~min there are clear
variations in the intensity producing an asymmetric fishbone pattern, as dark lanes propagate first to
one edge of the filament, then to the other. Somewhat weaker oscillations occur between 50 and
80 minutes, and these are followed by larger amplitude oscillations from $t\approx 100$~min onwards.

These bolometric intensity variations correspond to variations
in the vertical velocity, with the minima in intensity corresponding to stronger downflows.
These fluctuations are accompanied by variations in the $y$ component
of the velocity, corresponding to the tube at this height swaying first in the negative $y$
direction, and then in the positive $y$ direction. The second, weaker burst of oscillations
occurs between $t=50$ and $t=90$ min. The intensity fluctuations associated with this second
set of oscillations are more pronounced for $y<700$~km. The weak velocity fluctuations,
again best seen in $v_y$, indicate a swaying of the tube in the $y$ direction, consistent with magnetic field
strength variations which are asymmetric. The period of the oscillations in both phases is about 8 min.
We emphasize that the perturbations occur across the entire inhomogeneous filament,  despite the
tube having very different velocities at different locations.

The vertical structure of $v_y$, as a function of time at the three points indicated by stars in Fig.~\ref{fig:I_times}
is shown in Fig.~\ref{fig:vz_vert_slice}. At $x=25.7$~Mm, $y=510$~km (top frame of Fig.~\ref{fig:vz_vert_slice}), we see that mostly $v_y<0$,
corresponding to an outflow away from the central axis of the filament. At $x=25.7$~Mm, $y=660$~km (middle panel),
we see that along the center of the filament neither flows in the positive or negative
$y$ direction dominate and the clearest signature is of oscillatory motions. The situation
at $x=24.5$~Mm, $y=810$~km (bottom panel of Fig. 5) is conceptually similar to that at $y=510$~km, except that
on this side of the filament $v_y>0$ mostly dominates  which again corresponds to a lateral outflow
from the filament.  Oscillatory motions can be seen, and are in phase
at $y=510$~km and $y=660$~km, and especially at early times at $y=810$~km. The oscillations are
mainly propagating downwards (towards lower $z$ at later times). We measured the
wavelength to be $\approx 730$~km.

As well as studying the $x$ dependence of the oscillations at a particular time as in  Fig.~\ref{fig:I_times},
we also took a space-time  cut along the violet line in  Fig.~\ref{fig:I_times}. This
cut is shown in Fig.~\ref{fig:iisl}. The red line is placed at 25.7~Mm, corresponding to the red line in
Fig.~\ref{fig:I_times}. Oscillations can be seen near the red line between $t=0$ and $t=30$~min and
between $t=50$ and $t=90$~min. They appear as light and dark ridges running from the umbral end
of the filament towards the granulation. The apparent $x$ component of the wavelength is on the order of
1Mm.

The phase speed of the oscillations $\omega/\sqrt{k_x^2+k_z^2}$ is then approximately 2~km/s.
The wavevector, $(k_x,0,k_z)$, is inclined by approximately 45$^{\circ}$ to the vertical, directed downwards and
away from the umbra. The latter is consistent with the fact that the striations, when observed near the limb,
appear to propagate only away from the umbra.

To visualize the waves, Fig.~\ref{fig:T_cuts} shows the temperature in vertical cuts through the filament at the
location indicated by
the red line in Fig~\ref{fig:I_times} at $t=11$ and $t=16.5$~min, corresponding to two
nearly opposite phases of the oscillations. The differences in the temperature structure at the two phases
is large in the top 200~km of the filament, indicating that the oscillations are outside
the linear regime.

We comment that the mode is global with respect to the penumbral filament, although the filament has strong
velocity, temperature and field gradients. For this reason we think it is dangerous to interpret the associated
intensity fluctuations, such as those plotted in the top left of  Fig.~\ref{fig:iisl},
as simple tracers of the velocity field.

There are numerous physical forces and processes which affect the oscillations. The period of 8 minutes
(similar to the lifetime of granules) is long enough to make radiative processes important, the magnetic
field is strong and structured, the flows are a significant fraction of the local sound speed and highly
structured, whilst the density varies strongly across the filament and with depth. We leave the difficult
task of disentangling the various waves and instabilities which could play a role to a future study.

\subsection{Round sunspots}
To see if the oscillations found above are specific to penumbral filaments in the slab geometry, we looked
for similar features in the sunspot simulation described in \cite{Rempel2009b,Rempel2011}.
The simulation domain is periodic in both
horizontal directions ($x$ and $y$), with dimensions of 98.304~Mm $\times$
49.152~Mm, and has a vertical ($z$) extent of 6.144 Mm. The boundary
conditions are identical to the slab simulation described above. The setup of
this simulation contains a pair of opposite polarity sunspots with about
$1.6\cdot 10^{22}$~Mx each and maximum field strengths of about $3$ and $4$ kG, respectively. This setup leads to an
extended penumbra
in between both spots and we focus here our investigation on a part of the penumbra
belonging to the (left side) for the $4$ kG sunspot (see, e.g., Figs. 1 of \citet[see, e.g., Figs. 1 of]{Rempel2009b,Rempel2011}).
This part of the penumbra is the most extended
displaying long, well formed fibrils, which are shown in Fig.~\ref{fig:MS1}. A detailed
analysis of this region was recently performed by \citet{Rempel2011}.

\begin{figure}
\includegraphics[width=4cm]{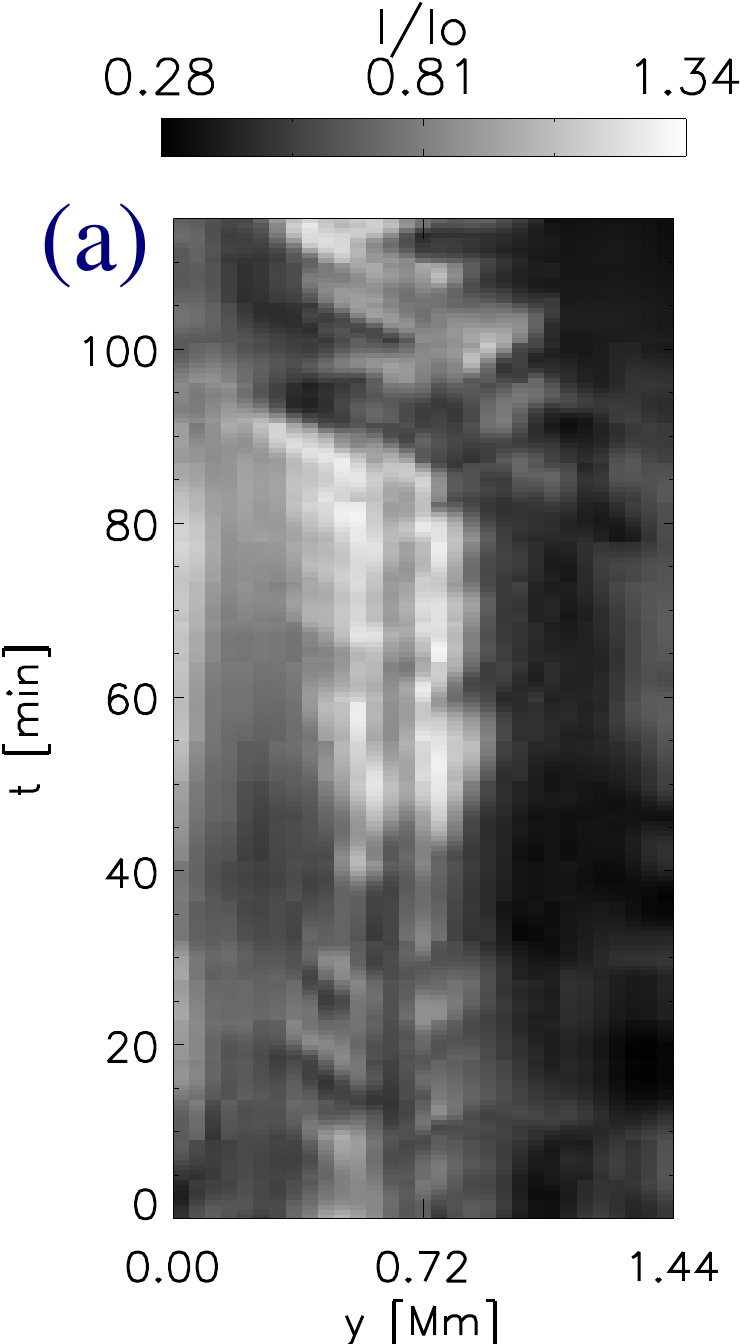}\includegraphics[width=4cm]{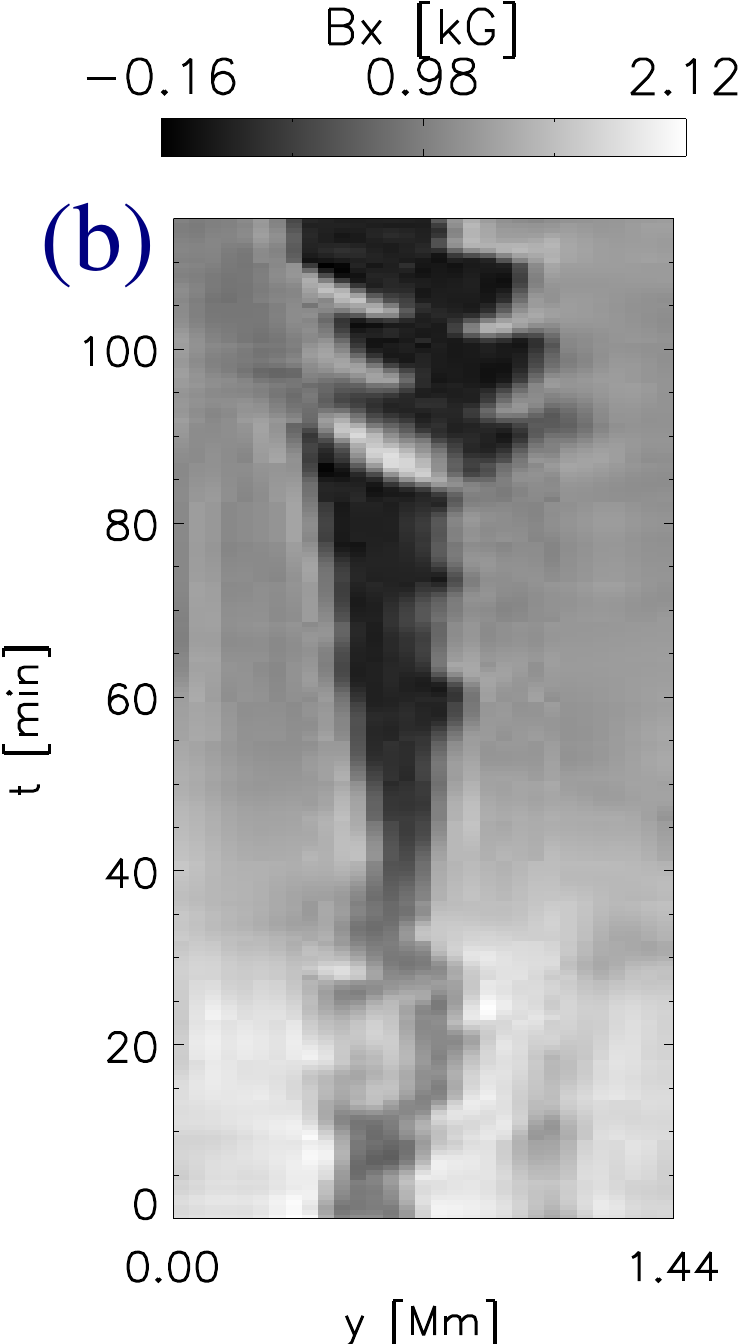}\includegraphics[width=4cm]{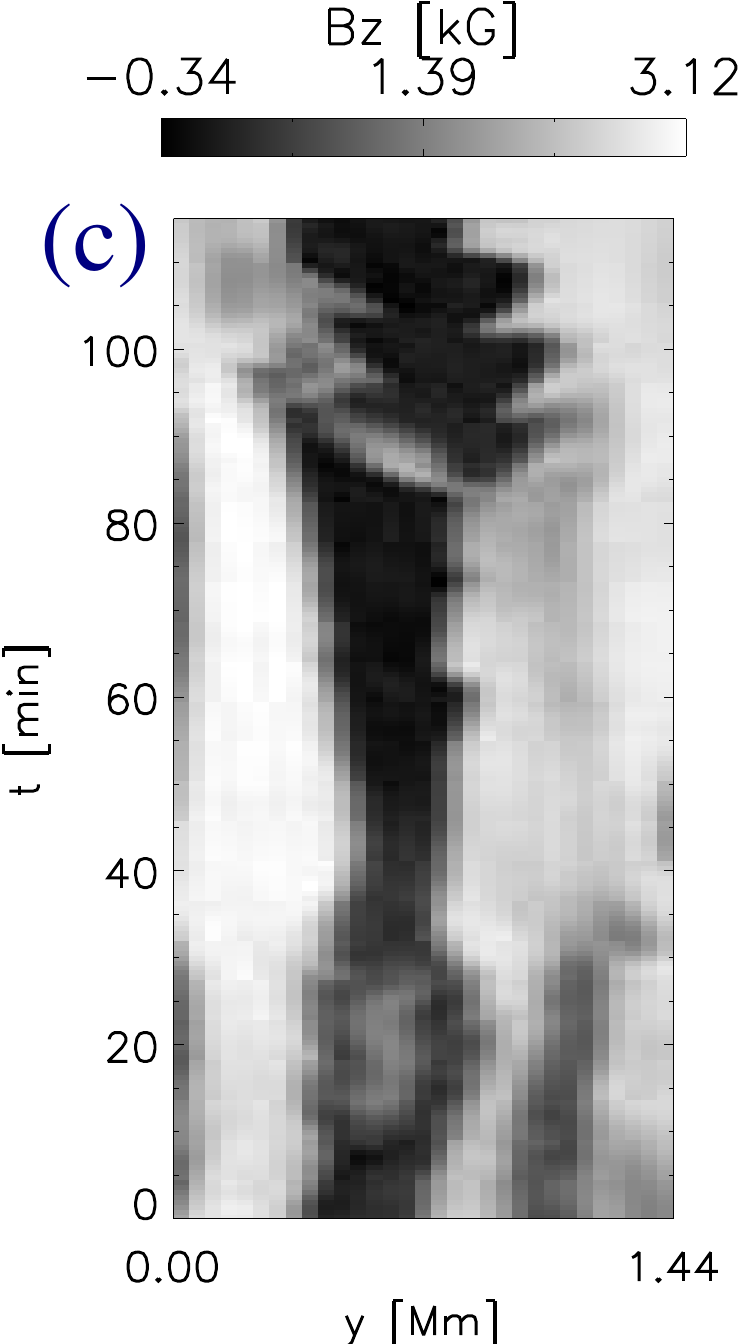}

\vspace{1.5cm}

\includegraphics[width=4cm]{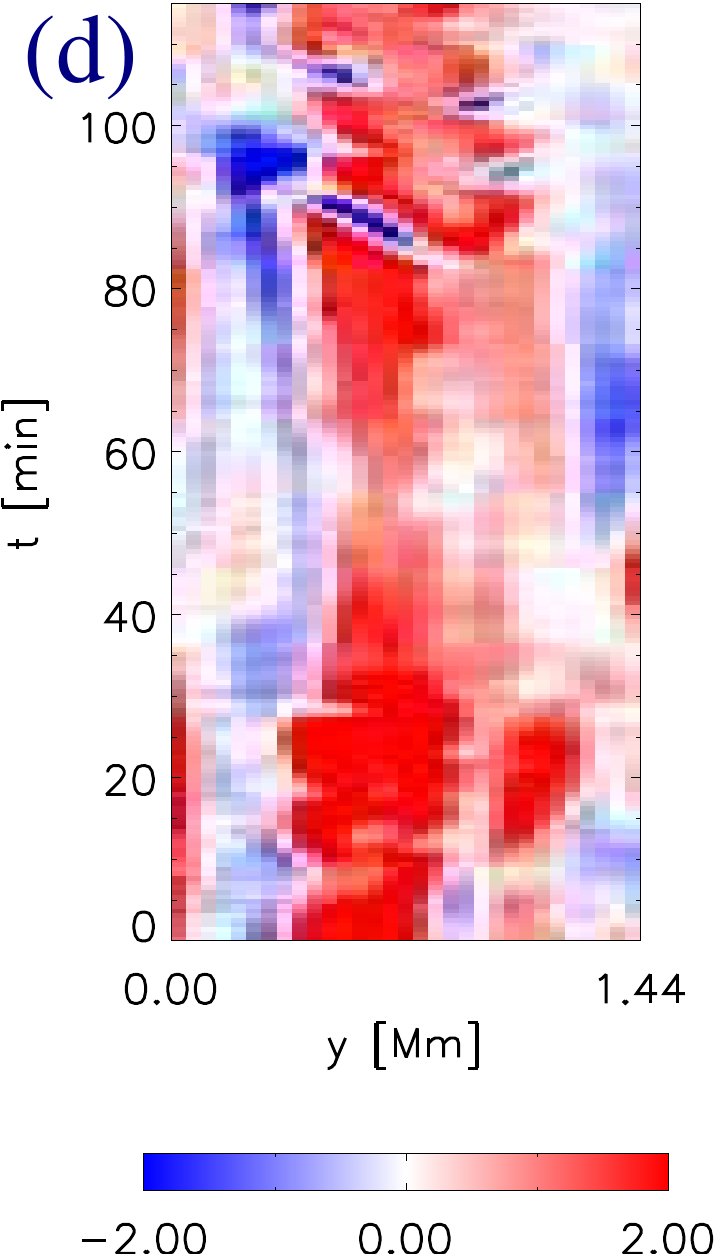}\includegraphics[width=4cm]{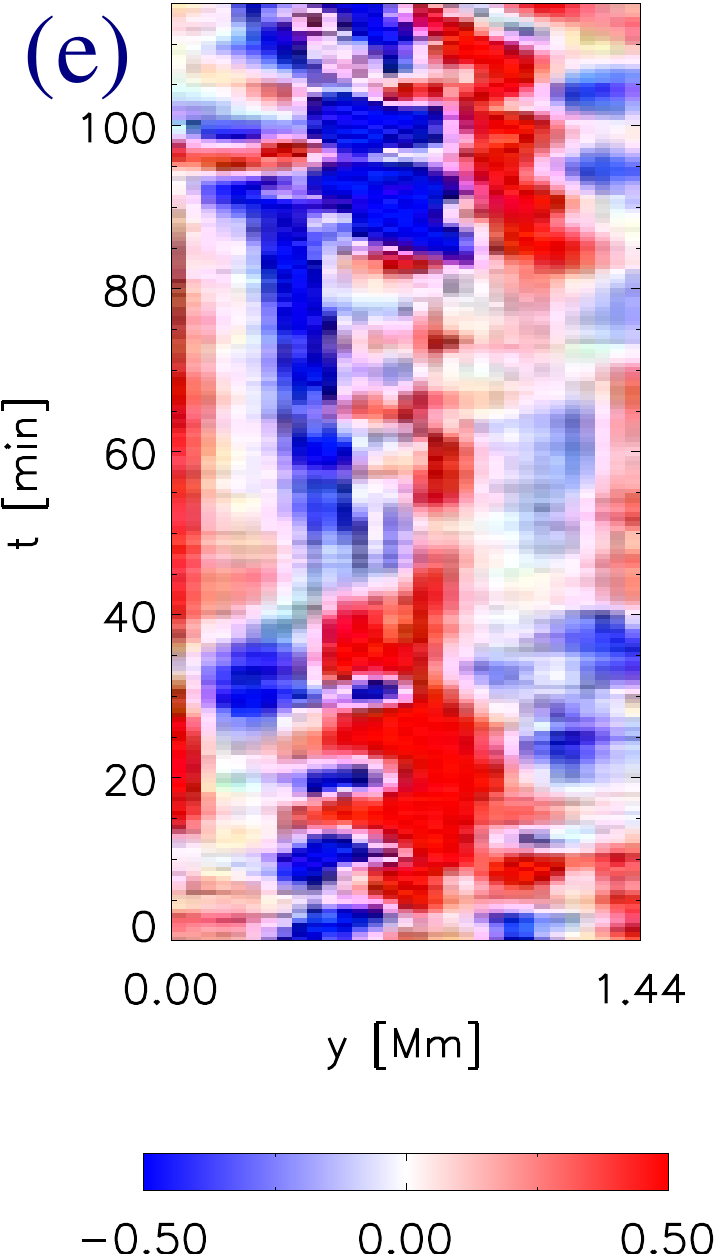}\includegraphics[width=4cm]{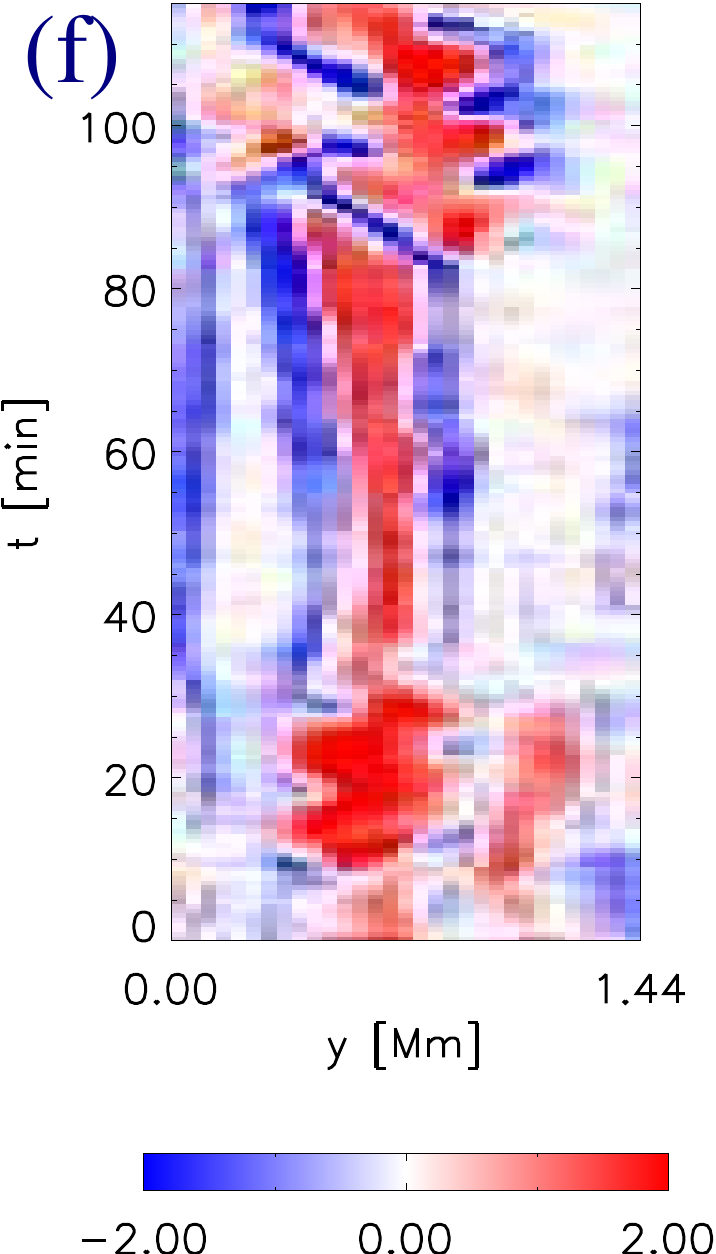}
\caption{Space-time plots from the MHD simulations at the slit location marked in red in Fig.~\ref{fig:I_times}(top panel).
Shown are (a) the bolometric intensity, (b) the $x$ component of the magnetic field,
(c) the vertical component of the magnetic field, (d) the $x$ component of the velocity, (e) the $y$ component of
the velocity, and (f) the vertical component of the velocity.
The  last five quantities are sampled at a constant geometric height of
$z=-384$~km below the average $\tau_{\mathrm Ross}$ height of the quiet Sun.}
\label{fig:I_time_slice}
\end{figure}

\begin{figure}
\includegraphics[width=17cm]{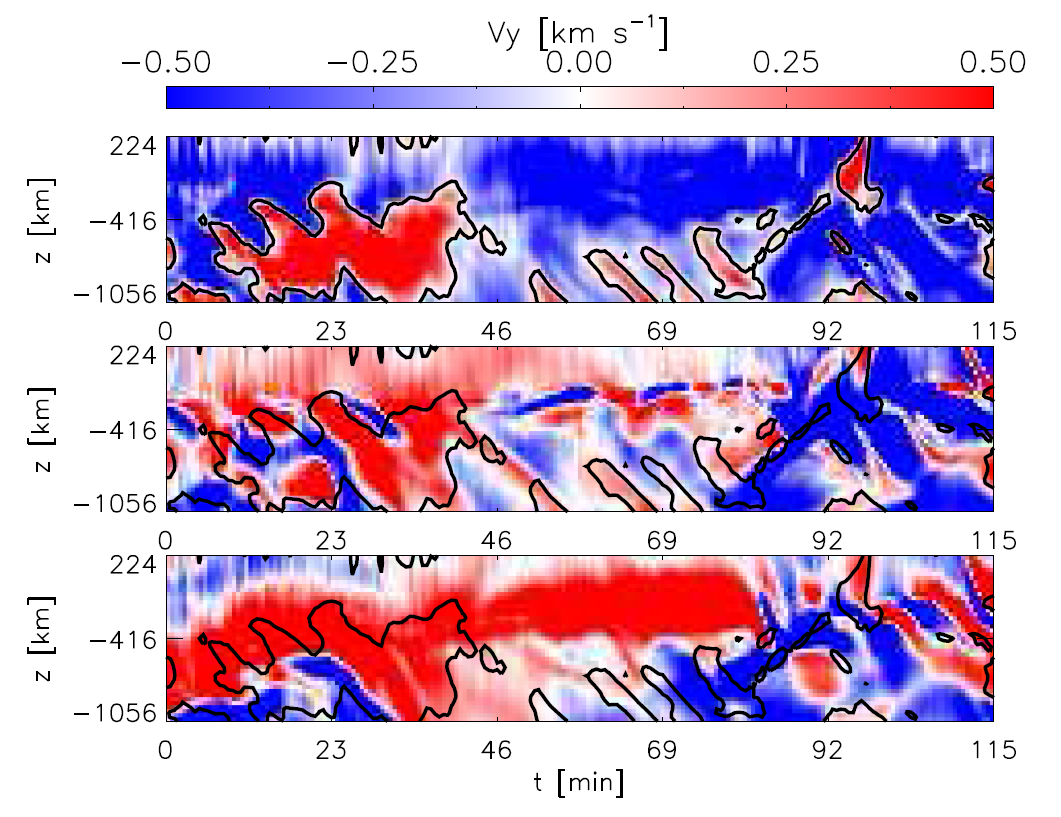}
\caption{Space-time plots from the MHD simulations at the three locations marked by crosses
along the slit marked in Fig.~\ref{fig:I_times}, i.e. from top to bottom at
$y=510$~km, $y=660$~km and $y=810$~km, respectively.
The spatial dimension corresponds to height, with $z=0$ corresponding to the average value of
$\tau_{\mathrm{Ross}}=1$ in the qiet Sun. Positive values of $z$ correspond to heights above the quiet-Sun $\tau=1$ level. The $y$ component of the velocity, saturated between $\pm 500$~m/s, is shown.
The black liness show where $v_y=0$ at $y=510$~km, they are intended to provide a reference for comparing the
structure of the velocity field between all three sub-images.  }
\label{fig:vz_vert_slice}
\end{figure}

\begin{figure}
\includegraphics[width=15cm]{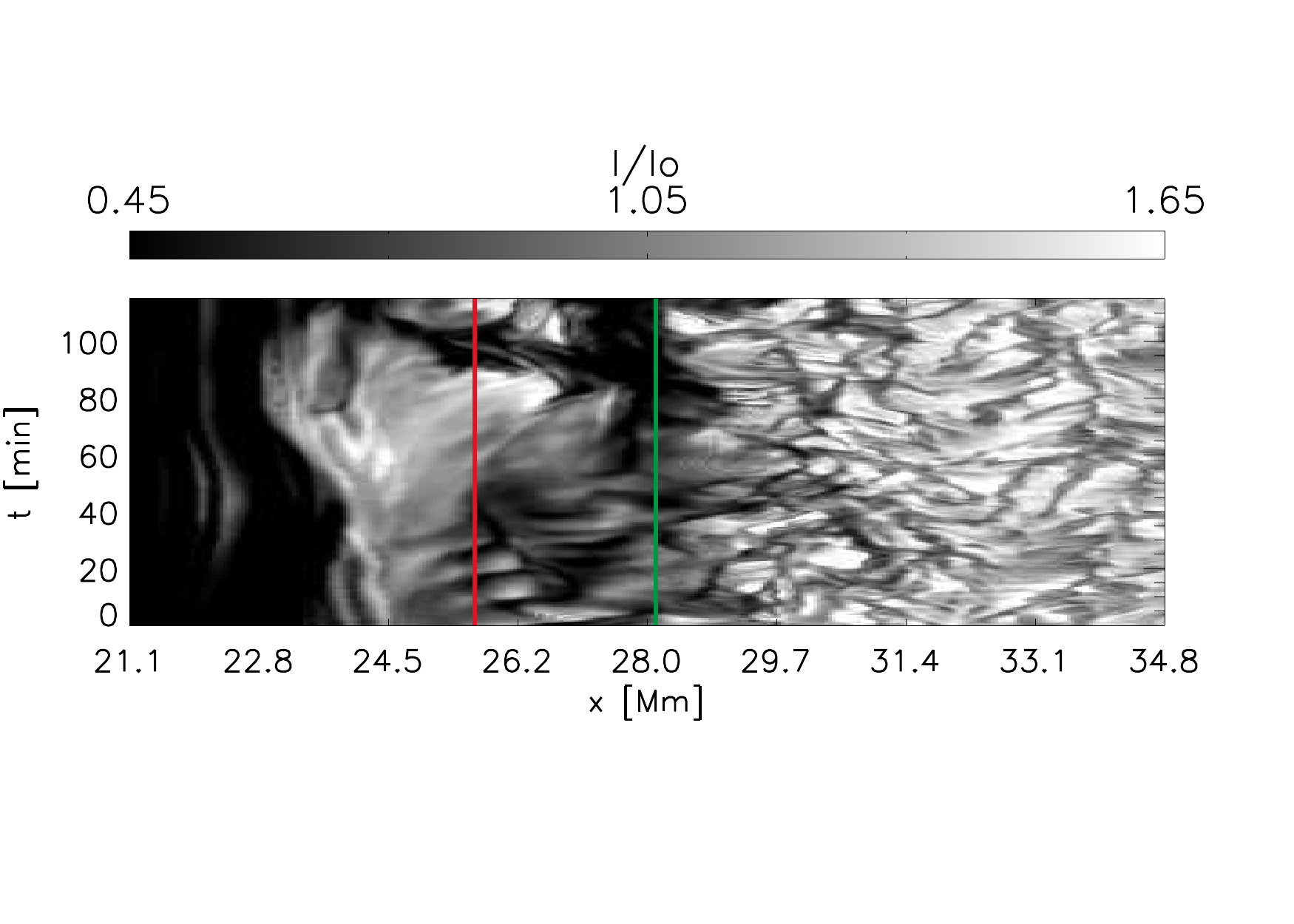}
\caption{Space-time plots of the normalized bolometric intensity along the blue line shown in
Fig.~\ref{fig:I_times} extended into the 'quiet-Sun' granulation. This cut is along the penumbral filament,
but offset from the centre of the filament.
The red line shows the $x$ value used to make Figures~\ref{fig:I_time_slice}, and \ref{fig:vz_vert_slice}.
To give an impression of how the signal discussed here differs from that of the umbra and quiet Sun,
we have used a bigger box than that used in Fig.~\ref{fig:I_times}, with the green line showing the extent
of  the box shown in Fig.~\ref{fig:I_times}.}
\label{fig:iisl}
\end{figure}

\begin{figure*}
\includegraphics[width=5cm]{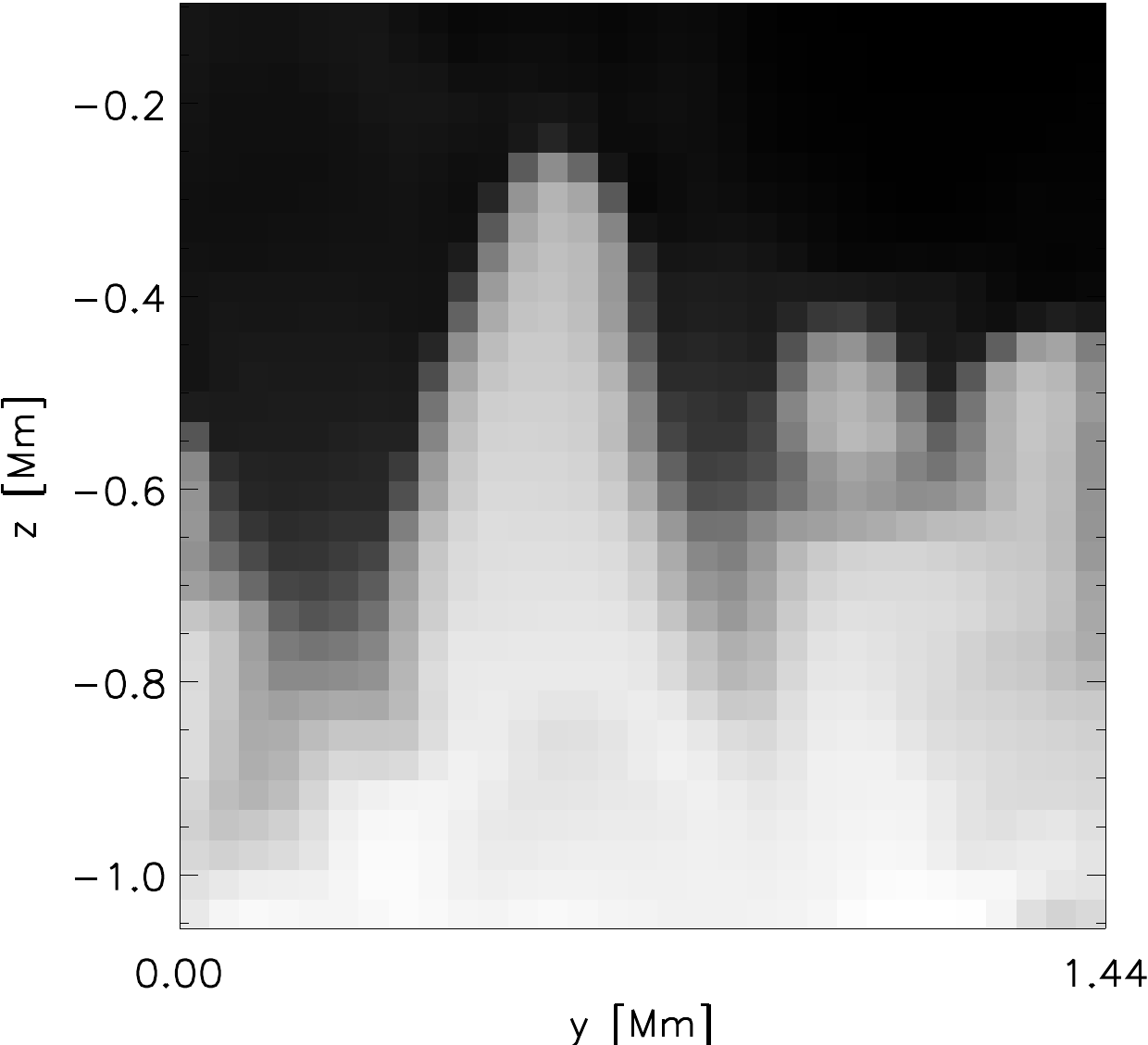}\includegraphics[width=5cm]{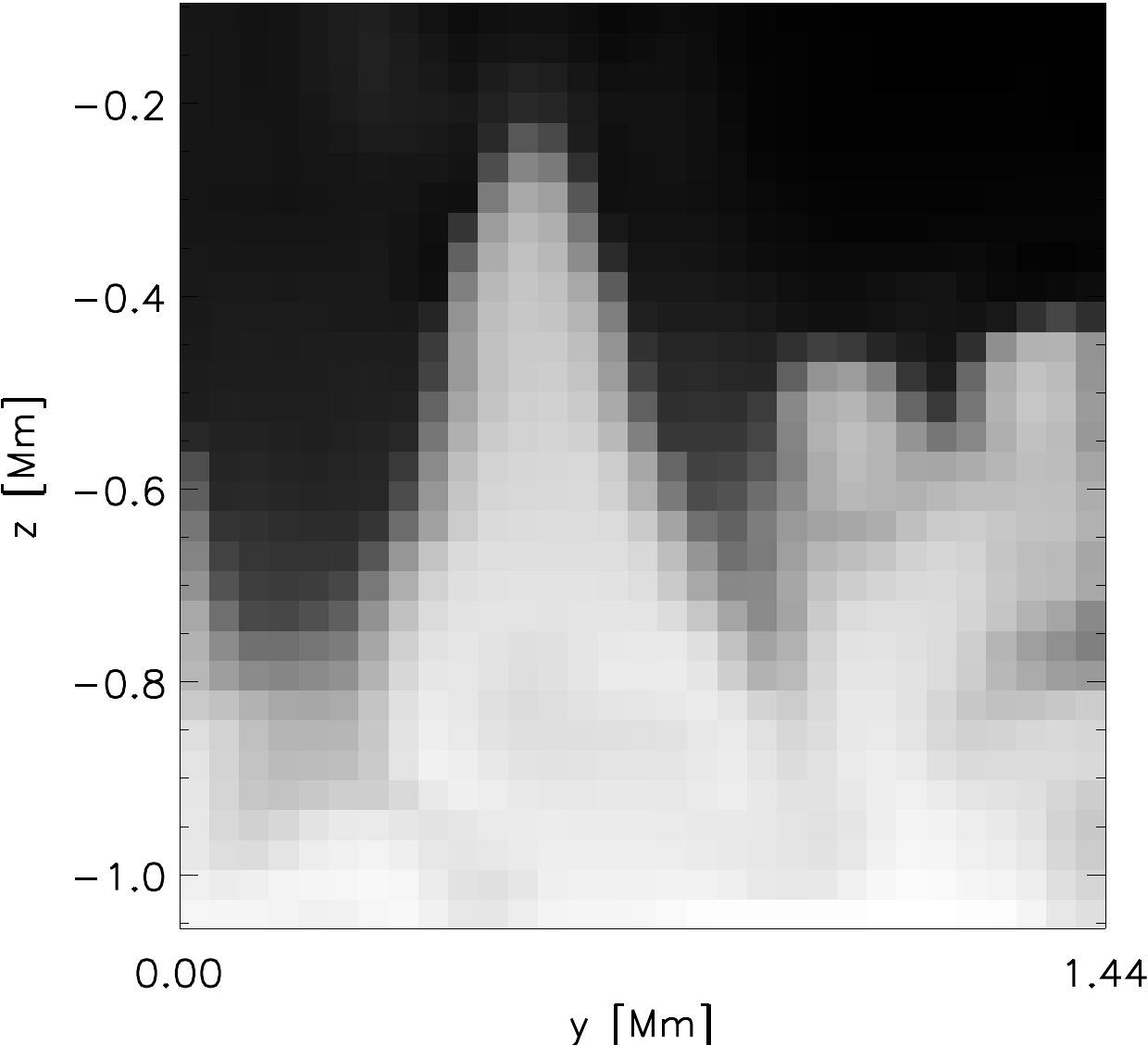}\includegraphics[width=0.93cm]{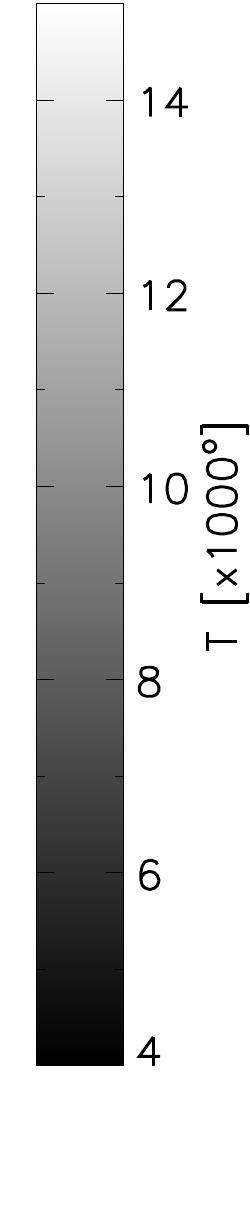}
\caption{Vertical cuts through the temperature field of the filament shown in Fig.~\ref{fig:I_times} at
different phases of the oscillation, $t=11$~min (left panel) and $t=16.7$ (right).
The average height of $\tau_{\mathrm Ross}=1$ in the quiet Sun was used to define $z=0$, with $z$ increasing outwards. }
\label{fig:T_cuts}
\end{figure*}

\begin{figure*}
\vspace{-10mm}
\includegraphics[width=12cm]{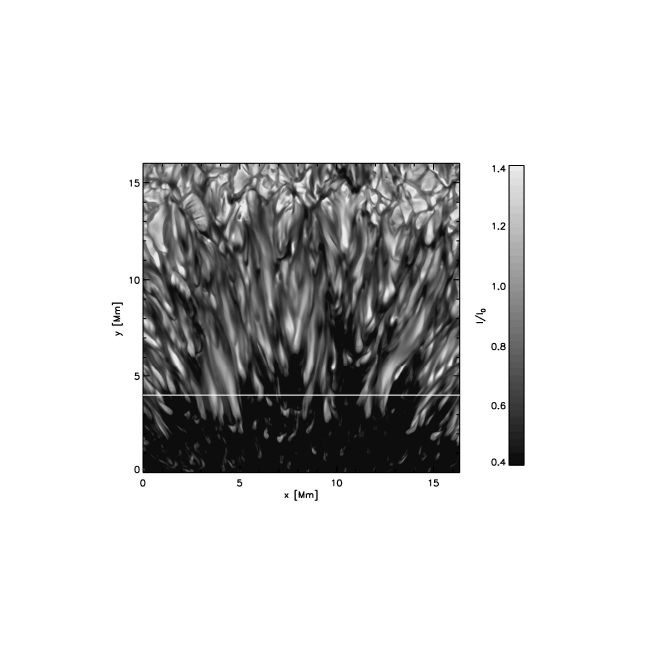}
\vspace{-20mm}
\caption{ Bolometric intensity image of the penumbral section analyzed from the simulation of
\cite{Rempel2009b, Rempel2011}. The white line indicates the slit position used for the time-slices shown in
Fig.~\ref{fig:MS2}. An animation of this figure is available in the online journal.}
\label{fig:MS1}
\end{figure*}

\begin{figure*}
\includegraphics[width=15cm]{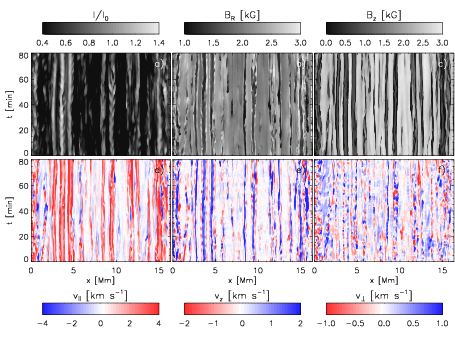}
\caption{ Space-time plots from the MHD simulation shown in Fig.~\ref{fig:MS1}.
Shown are a) the bolometric intensity, b) strength of the magnetic field component in the radial direction (with respect
to the approximate center of the spot), c) strength of the vertical magnetic field component, d) velocity component along the magnetic field's
direction (in the horizontal plane), e) vertical flow velocity, and f) velocity component perpendicular to magnetic field direction in the horizontal plane. Magnetic field and velocity data are for a constant geometric
height $z=-300$~km below the average $\tau_{\mathrm Ross}$ height of the quiet Sun.}
\label{fig:MS2}
\end{figure*}

\begin{figure*}
\includegraphics[width=15cm]{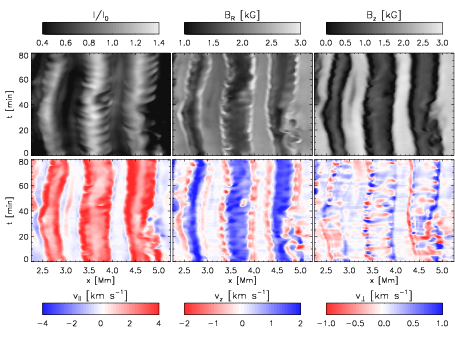}
\caption{Same quantities as in Fig.~\ref{fig:MS2} for the filaments located between $x=2.25$ and $x=5.25$~Mm.}
\label{fig:MS3}
\end{figure*}

\begin{figure*}
\includegraphics[width=15cm]{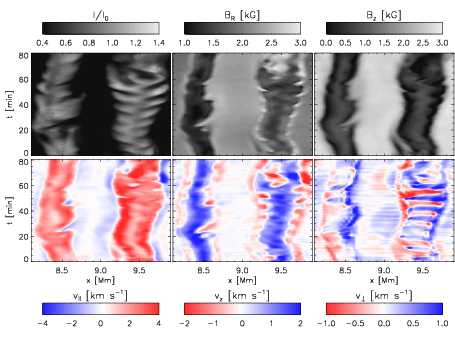}
\caption{Same quantities as in Fig.~\ref{fig:MS2} for the filaments located inbetween $x=8.1$ and $x=9.9$~Mm.}
\label{fig:MS4}
\end{figure*}

The spatial location of the space-time slices plotted in Fig.~\ref{fig:MS2} is marked by the horizontal line in
Fig.~\ref{fig:MS1}. The time slices displayed in Fig.~\ref{fig:MS2} are computed using a constant geometrical
height about $300$~km beneath the quiet Sun $\tau=1$ level, which is below the local $\tau=1$ in bright filaments
and above it in the dark filaments. The presence of oscillations in this simulation is suggested by the typical
asymmetric fishbone pattern, earlier detected in the slab simulation, most clearly visible in the filament at the
position $x=9$~Mm. This suspicion is heightened by the clear oscillatory signal in the velocity perpendicular
to the filament. A more detailed view of the filaments located between $x=2.25$ and $x=5.25$~Mm as well
as $x=8.1$ and $x=9.9$~Mm is shown in Figs.~\ref{fig:MS3} and \ref{fig:MS4}, respectively.
Clearly, this simulation  also displays oscillations in
several filaments in the innermost parts of the penumbra with periods around 7-8 minutes. The most
significant difference compared to the oscillations in the slab geometry is a substantially smaller horizontal
displacement. While the slab simulation presented in Fig.~\ref{fig:I_time_slice} shows a lateral
displacement of the filament with an amplitude comparable to the width of the filament in all plotted variables, the
oscillations here occur within otherwise mostly unaffected flow channels. The
oscillation is most prominent in the intensity and in the velocity component lateral to the filament. Variations in
the magnetic field strength are mostly restricted to a narrow boundary layer characterized by enhanced horizontal
and reduced vertical field strength. This boundary layer coincides with the region of convective downflows at the
edge of the filaments, as has been shown in \citet[][see Fig. 17 therein]{Rempel2011}. The vertical flow velocity
shows moderate changes in the central upflows, but a rather intermittent behavior in the lateral downflows. The
average outflow velocity from filament has its largest amplitudes near the outer edge of the filaments, where we also find
the strongest flow variations. They remain small compared to the mean flow velocity, however. Both, the radial magnetic
field and radial outflow originate mostly from a thin boundary layer just below the $\tau=1$ level
\citep[see Fig. 17 of ][]{Rempel2011}, where strong horizontal field is induced and the Evershed flow is driven. The
horizontal field and accelerated fluid are transported downward by the overturning convection within the filament,
leading in deeper layers to a filament structure with enhancements of both radial magnetic field strength and flow
velocity at the lateral boundaries of the filaments. The most significant difference between this simulation and the
previously discussed slab simulation is in terms of a
substantially stronger Evershed flow that is accompanied with stronger horizontal field within
flow channels: the flow velocity along filaments reaches in most filaments 3 km/s, while the horizontal field
strength remains of the order of 1.5 kG. The horizontal field enhances the stiffness in the direction along
filaments and thus suppresses lateral deformations. This could be taken as an indication that the
lateral displacement of the flow channel in Fig.~\ref{fig:I_time_slice} is more a consequence than a
cause of the oscillation mode. Furthermore the substantial difference in flow velocity and field strength
does not seem to affect the oscillation period compared to the example shown in Fig.~\ref{fig:I_time_slice}.

\section{Observations}

For the present study we selected a time series of G-band (4305\,\AA ) images
of a sunspot located almost at disk center and recorded by the Broadband Filter Imager (BFI) of the
Solar Optical Telescope (SOT)
onboard {\it Hinode} \citep{tsuneta07}. Such a location differs significantly from all previous studies of moving dark stripes in penumbral filaments, which were all carried out closer to the limb. The time series, recorded on
January 5, 2007, consists of 432 images at 30\,s cadence and contains a
sunspot (NOAA 10933) located at disk center. The spatial
resolution in the G-band is approximately 0\farcs22. The image scale is
0\farcs054 per pixel. The Solar Soft pipelines for the Hinode SOT/BFI were used for flat field and dark
current corrections.  The images are reconstructed for the instrumental PSF, applying a Wiener filter \citep{sobotka93} and assuming diffraction on an ideal circular 50 cm aperture.
Finally, a subsonic filter \citep{title89} with a cut-off velocity of 6\,km\,s$^{-1}$ is used to filter out contributions
of five-minute oscillations. The first and last 14 images from the time series of January 5, 2007 have been omitted due to the
apodizing window used in the subsonic filtering. The mean intensity of quiet regions (i.e., regions containing undisturbed
granulation and an absence of bright points) close to the observed sunspot were used for intensity normalization in all images.

\subsection{Observational analysis}

Figure 12 shows a G-band image of the sunspot at the beginning of the filtered time series.
A central umbra and a fully developed penumbra can be seen.
Filaments oriented in all azimuthal directions clearly exhibit central dark cores. An animation
of the analyzed time series reveals a lateral motion of dark lanes in the
filaments from the axis of a particular filament toward either one of its edges, or toward both of its edges.
These lateral motions are found in all azimuthal directions from the center of
the sunspot. They are strongest at the inner ends of the filaments, in
particular in filaments that protrude into the umbra.

Two representative space-time diagrams are shown in Figs. 13 and 14, respectively.
they correspond to the lines 'S1' and 'S2' lying on opposite sides of the penumbra in Fig.
12. The motion of thin dark stripes across some of the filaments is apparent as inclined
dark stripes in Figs. 13 and 14.

Figure 13 shows the space time diagram for the horizontal slit 'S1'.
The white boxes indicate locations of filaments where inclined stripes or "twists" can be clearly seen.
Only some of the filaments show such stripes and only for a part of the time.
In the panels on the right, each box is displayed two times, once simply to highlight
the stripes, the second time with white lines overplotted on the dark stripes for better visibility.
Filament 'A', located at $x=19$\farcs5, shows a central dark core and dark stripes,
from 0 to 20 min, pointing toward both edges.
The dark stripes in opposite directions
occur alternatively, recalling from the spacing between asymmetric fishbone appearance already seen in the left panel of Fig. 4.
This pattern can also be seen for filament 'B' at $x=23$\farcs5 from
22 to 38 min.  Filament 'C'  at $x=27$\farcs0 shows stripes in one direction only. Filament 'D'
located at $x=23$\farcs0 shows dark stripes from 100 to 135 min also just in one direction, but
opposite to 'C'. Such "one-sided" stripes may correspond to the weaker,
asymmetric oscillations seen between 50 and 90 min in the slab simulations (Fig. 3) and the leftmost
filament located between 2.5 to 3.0 Mm in Fig. 9 for the round spot.

Figure 14 is a space-time plot along  slit 'S2' depicted in Fig. 12. Filament 'E', located
at $x=27$\farcs0, shows stripes in one direction from 30 to 75 min.
Filament 'F' at $x=29$\farcs5 shows an asymmetric fishbone pattern from
56 to 70 min. Figure 14 demonstrates that such stripes are not restricted to a single location and that,
because the sunspot is nearly at disk center, opposite sides of the penumbra display essentially the
same behavior (in a statistical sense).
From these two images we estimate that the period of the observed oscillations is in the range of 3 to 7 minutes.

Space-time diagrams along  further lines (slits) cutting across filaments slits oriented perpendicular to S1 and S2 (e.g. at solar $x=-22$ and at +4 arc sec)
were also produced (not shown). Similar stripes and patterns as shown in Figs. 13 and 14 can also be recognized there,
although slightly less clearly.

It is also instructive to consider space-time diagrams  running along the filaments (i.e. along slit S3 in Fig. 12), the observational analogy
to that shown for the simulation in Fig.~\ref{fig:iisl}. The oscillations we saw in Figs. 13 and 14 are also visible in Fig.~15.
They can be clearly seen between $y=-35$ and $y=-34$ ~arsec between $t=0$ and $t=30$ as a stack of dark and white stripes, whose
inclination to the horizontal indicates that the waves are propagating outwards from the umbra. However, it is also clear that
oscillations and propagating waves are ubiquitous. In particular, on the opposite side of the umbra waves (now propagating in the
opposite direction, i.e. still away from the umbra) are visible at many locations. From this figure we can also determine that the
wavelength along the filament is about 500 km.

Not only are moving stripes visible also at
disk center and not just at the limb, but they also move with roughly the same phase-speed (1.2-2.8\,km\,s$^{-1}$)
and display roughly the same periodicity (3-7 min) at disk center as they do closer to the limb.
This observed velocity is consistent with the apparent phase velocity inferred from the simulation,
e.g. 2\,km\,s$^{-1}$ for the slab geometry (see Section~2.1.1), although the period is somewhat
larger in the simulations (possibly related to the subsonic filtering we applied to the observations),
e.g. 8~min in the slab geometry and 7-8~min for the round sunspots.

\section{Conclusion}

We analyzed striations moving across bright penumbral filaments  in both numerical simulations and in a time series of seeing-free G-band images obtained with
{\it Hinode}. In both cases we find that these moving striations, which give the filaments a twisting appearance, are also visible at disk center
and appear relatively similar in simulations and observations. Interestingly, at disc center the two halves of a filament on either side of its centeral dark core can display "twists" in opposite directions, with the stripes on the two sides being out of phase.

Two different numerical simulations indicate that the striations are
oscillations propagating along the penumbral filaments directed away from the umbra and downward. The periods of the oscillations one found to be 3-7 min from observations and 7-8 min in the simulations. The explanation for the striations suggested by the simulations is quite different from that proposed by \cite{spruit10} and, in particular does not support their conclusion that the striations imply the absence of a dynamically significant horizontal magnetic field strength in bright penumbral filaments.  The MHD simulations reproduce the data rather well, although they have a rather dynamically significant horizontal field in the filaments. Although the Simulations with a stronger horizontal field (1500 G) in the filaments produce oscillations with a smaller amplitude, they are still consistent with the observations.
These oscillations are potentially a new seismic diagnostic which can be used to better understand
penumbral filaments.  This would, however, require a physical understanding of the underlying oscillatory mode, which
itself will require further study.

\vspace{5cm}

\acknowledgments{
This work has been partly supported by the WCU grant No. R31-10016 funded by
the Korean Ministry of Education, Science and Technology.
Hinode is a Japanese mission developed and launched by ISAS/JAXA,
collaborating with NAOJ as a domestic partner, NASA and STFC (UK) as
international partners. Scientific operation of the Hinode mission is
conducted by the Hinode science team organized at ISAS/JAXA.
Support for the post-launch operation is provided by JAXA
and NAOJ (Japan), STFC (U.K.), NASA (U.S.A.), ESA, and NSC (Norway).
The National Center for Atmospheric Research in sponsored by the National
Science Foundation. Computing time was provided by NCAR's Computational and
Information Systems Laboratory (CISL).}

\bibliography{MURaM}

\newpage

\vspace{-1.5cm}

\begin{figure}
\vspace{-10mm}

\includegraphics[width=192mm,angle=0]{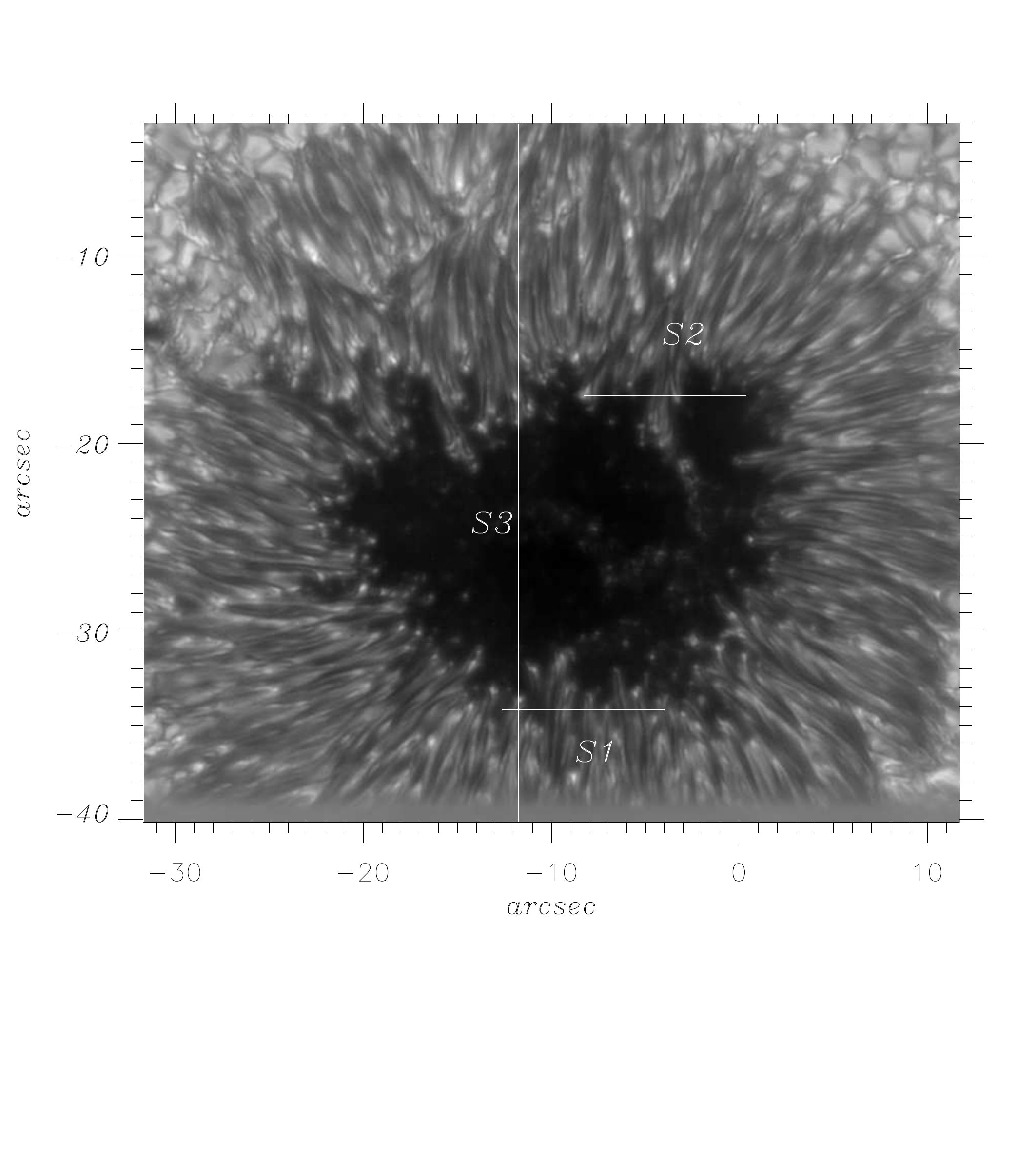}
\vspace{-35mm}

\caption{A sunspot image in the G-band taken on January 5, 2007 by the SOT/BFI aboard Hinode at disk center.
'S1' and 'S2'  point to the locations of
horizontal slits for which space-time diagrams have been made (see Figs. 13 and 15). Similarly 'S3' represents a vertical slit along which the space-time diagram displayed in Fig. 14 is constructed. Coordinate (0,0) corresponds to solar disc center.}
\end{figure}


\begin{figure*}
\vspace{-35mm}
\hspace{10mm}
\includegraphics[width=180mm,angle=0]{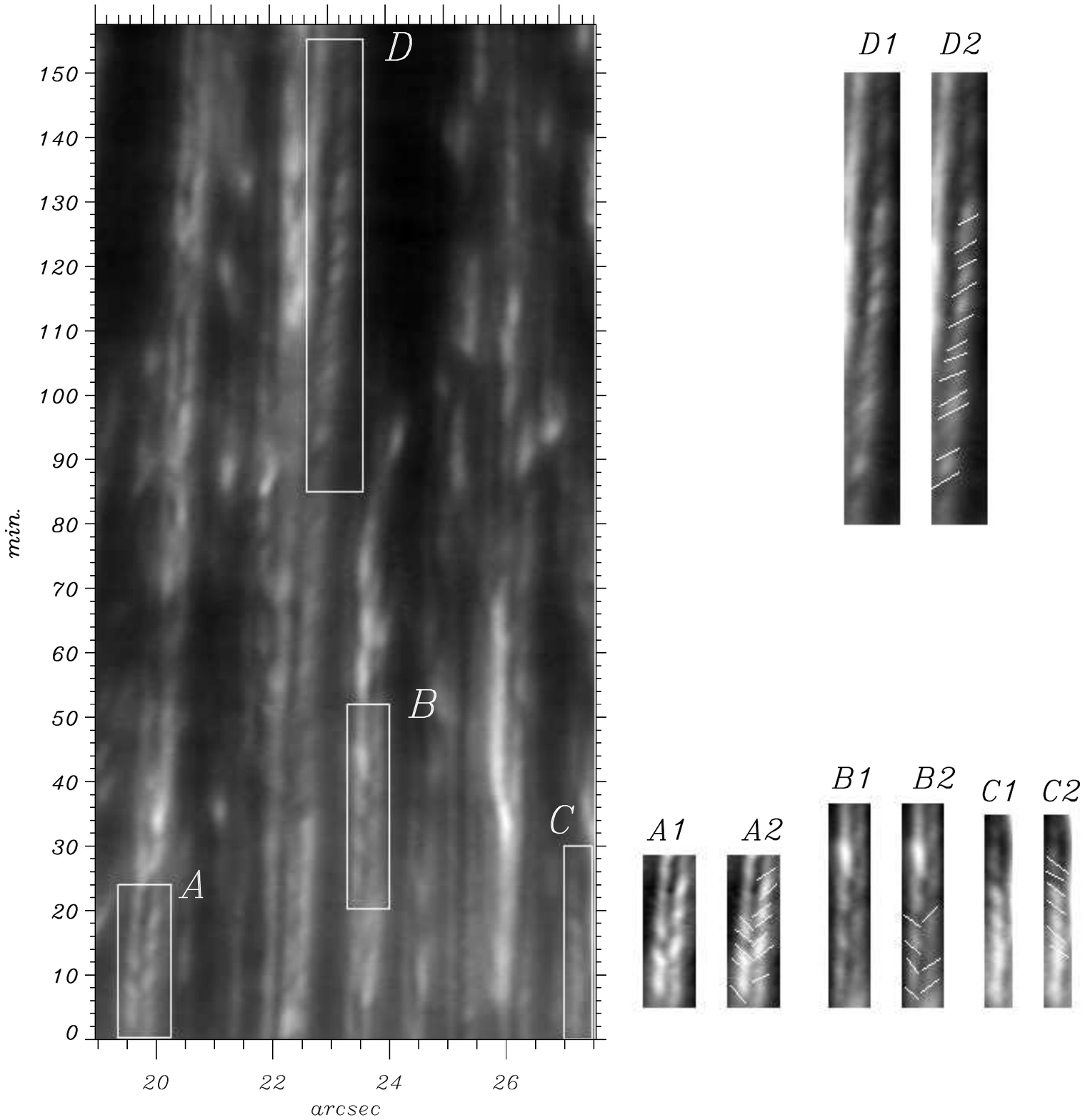}
\vspace{-35mm}
\caption{Large panel at left - Space-time slice along the line marked by 'S1' in Fig. 12. The highlighted boxes A, B, C and D are repeated in the small panels
A1, B1, C1, D1. Panels A2, B2, C2, D2 repeat panels A1, B1, C1, D1, but with white lines overlying the dark stripes to better reveal their slopes, which are
indicative of an apparent phase velocity.}
\end{figure*}

\begin{figure*}
\vspace{-35mm}
\hspace{15mm}
\includegraphics[width=180mm,angle=0]{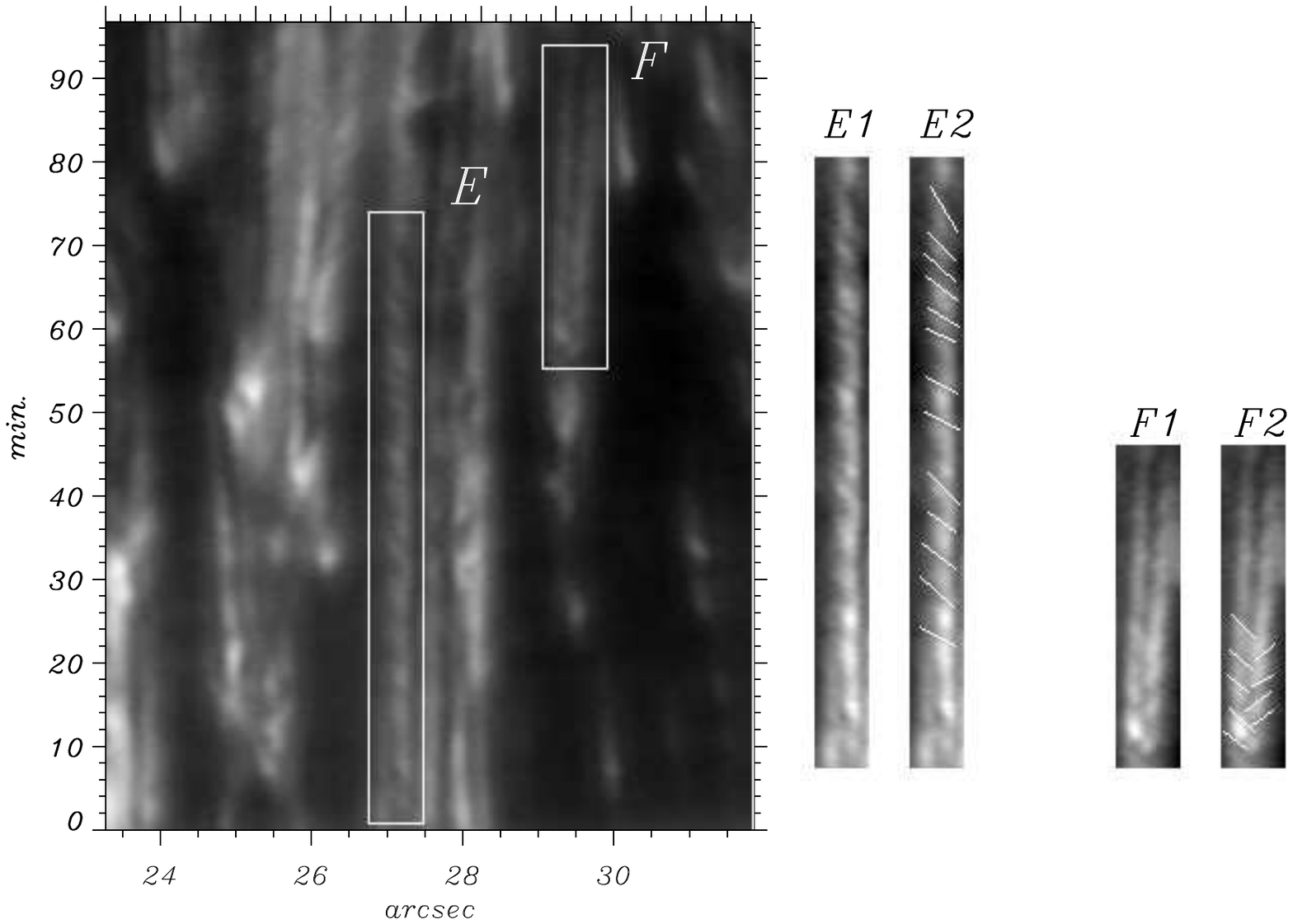}
\vspace{-35mm}
\caption{Same as Fig. 13, but along the line marked by 'S2' in Fig. 12.}
\end{figure*}

\begin{figure*}
\vspace{-10mm}
\hspace{-5mm}
\includegraphics[width=180mm,angle=0]{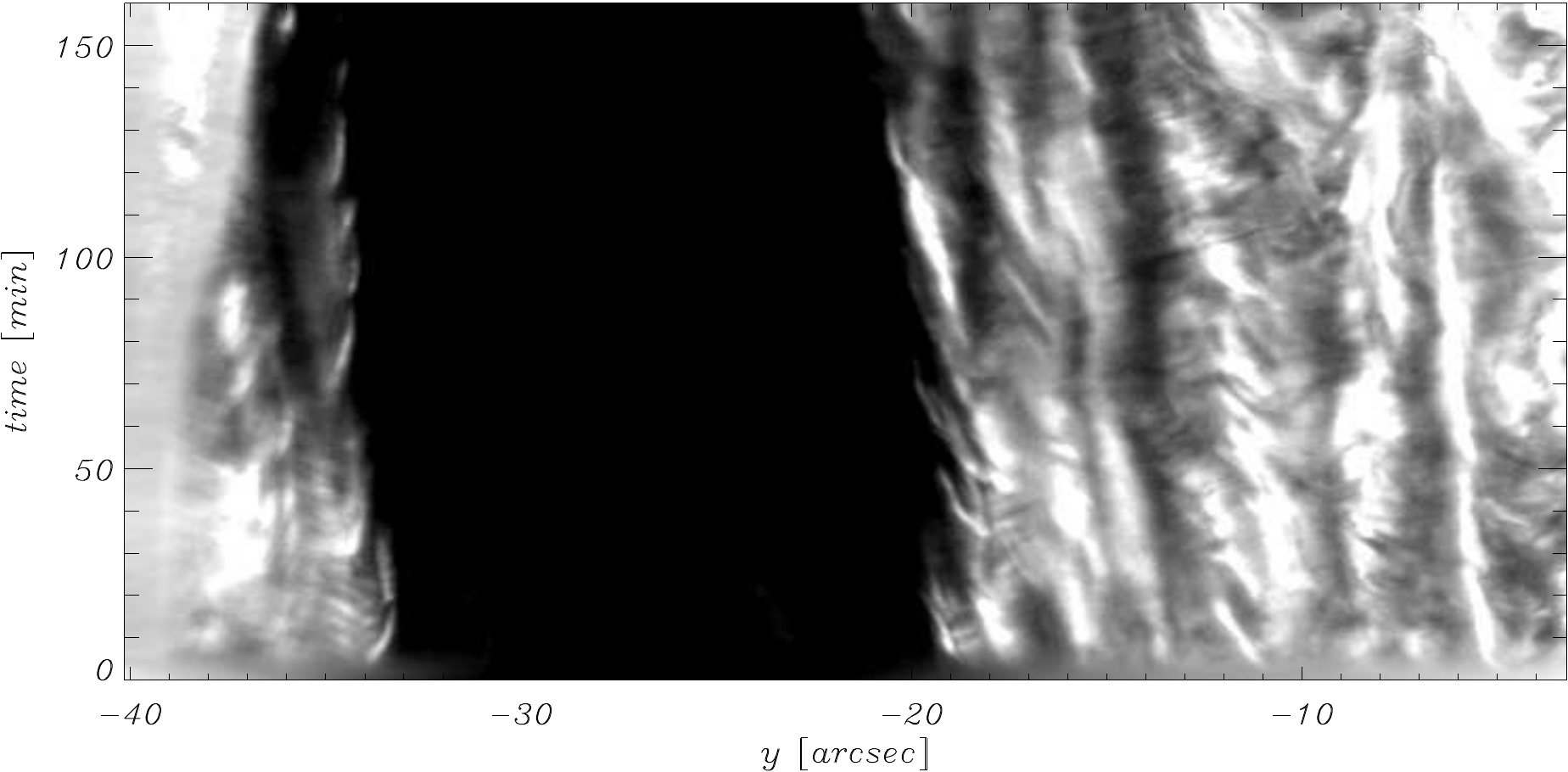}
\vspace{10mm}
\caption{Space-time slice along the line marked by 'S3' in Fig. 12, almost along the
filaments. The oscillotions seen at $y=-35$~arsec between $t=0$ and $t=30$
min are discussed in the main text.}
\end{figure*}

\end{document}